\documentstyle[12pt]{article}
\advance\hoffset by -1.1truecm

\begin{document}

\title{q-Oscillators, q-Epsilon Tensor, q-Groups}

\author{Metin Arik, Gokhan Unel \\ Department of Physics \\ Bogazici
University \\ Bebek 80815 Istanbul \\ TURKEY \\ \\ and
\\ \\ Muhittin Mungan \\Enrico Fermi Institute and Department of Physics\\
The University of Chicago\\ Chicago, IL 60637 \\ USA}

\date{August 2 1993}

\maketitle

\vskip-15cm
\hskip4.5cm
\tt preprint BUFB 93-05 \quad EFI 93-45 \quad hep-th/9308096 \rm
\vskip14.5cm

\begin{abstract}

	Considering a multi-dimensional $q$-oscillator invariant under the
(non quantum) group $U(n)$, we construct a $q$-deformed Levi-Civita epsilon
tensor from the inner product states.  The invariance of this $q$-epsilon
tensor is shown to yield the quantum group $SL_{q}(n)$ and establishes the
relationship of the $U(n)$ invariant $q$-oscillator to quantum groups and
quantum group related oscillators.  Furthermore the $q$-epsilon tensor
provides the connection between $SL_{q}(n)$ and the volume element of the
quantum hyper plane.

\end{abstract}

	With the discovery of the quantum groups \cite{[1]},
$q$-generalizations of the harmonic oscillator have recently been the center
of attention \cite{[2]}.  There exist different formulations of these
generalizations, the so called $q$ -oscillators.  Strict constraints on
these generalizations come from the multidimensionality of the oscillator.
One aspect that has been considered is invariance under the unitary quantum
group \cite{[3]}.  A related aspect that yields quantum group invariance is
the degeneracy \cite{[4]} of the eigenstates of the generalized bilinear
hamiltonian.

	The earliest multidimensional $q$-oscillator was postulated by
Coon, Baker and Yu such that its Euclidean version respects the $U(n)$
symmetry of the ordinary $n$ -dimensional oscillator.  It is defined by the
following relations\cite{[5]}:

\begin{eqnarray}
	a_i a^*_j &=& q a^*_j a_i + \delta_{ij}, \nonumber \\
	 a_i |\rangle &=& 0 \mbox{\quad i,j = 1,2, ...  D,\quad}
\end{eqnarray}

where $q$ is a real parameter, $a_{i}$ is the annihilation operator, its
hermitian conjugate, $a^{*}_{i}$, is the creation operator and $| \rangle$
denotes the ground state which is taken to be normalized.  The quantum
states belonging to the oscillator are then obtained by applying the
creation operators on the ground state.  Thus for example,

\begin{equation}
	a^*_i a^*_j a^*_k | \rangle = | i j k \rangle
\end{equation}

and its hermitian conjugate bra vector is given by ,

$$\langle | a_k a_j a_i = \langle i j k|.$$

In agreement with the conventional terminology we call a state obtained by
applying $n$ creation operators on the ground state, an "$n$-particle"
state.  From (1) it follows that states with different number of particles
are orthogonal.  The inner products of the states with same number of
particles satisfy

\begin{eqnarray}
	\langle i | j \rangle &=& \delta^{i}_{j}\nonumber \\
	\langle i j | k m\rangle &=& \delta^{i}_{m}
	\langle j | m \rangle + q \delta^{i}_{m}
	\langle j | k \rangle
\end{eqnarray}

and therefore by induction :

\begin{eqnarray*}
	\langle i_1 i_2 i_3 ...  i_n | k_1 k_2 k_3 ...  k_n \rangle
 &=& \delta^{i_1}_{k_1} \langle i_2 i_3 ...  i_n | k_2 k_3 ...  k_n \rangle
    + q\delta^{i1}_{k2}\langle i_2 i_3 ...  i_n| k_1k_3 ...  k_n \rangle \\
 &+& q^2\delta^{i_1}_{k_3}\langle i_2 i_3 ...  i_n | k_1k_2 k_4 ...  .k_n
   \rangle \\
 &+&  ...  +q^{n-1} \langle i_2 i_3 ...  i_n | k_1k_2k_3 ...
    k_{n-1} \rangle
\\
\end{eqnarray*}

{}From now on we will call the inner product

\begin{equation}
	\langle i_1 ...  i_n | k_1 ...  k_n \rangle := N^{i_1i_2
	  ...  i_n}_{k_1k_2 ...  k_n} \left( n,q \right) = N^{i}_{k}\left(
	  n,q \right)
\end{equation}

In the above definition, $N$ is a function of the real parameter $q$ and
 the number of particles $(n)$ is directly involved.

 	It is important to note
 that the relations (1) which define the $U(n)$ invariant $q$-oscillator, do
 not contain any commutation relations involving only the annihilation
 operators or only the creation operators.  In fact the existence of a
 Hilbert space with positive definite inner product, for $|q| = 1$
 determines these commutation relations, and for $-1< q < 1$ implies that
 there can be no commutation-like relation between two annihilation
 (creation) operators.  Thus for example, we find that the two particles
 states $| 1 2 \rangle$ and $| 2 1 \rangle$ are orthogonal to all other
 states; whereas their norms and scalar products satisfy

\begin{eqnarray}
	\langle 1 2 | 1 2 \rangle = \langle 2 1 | 2 1 \rangle &=& 1
	\nonumber \\
	 \langle 1 2 | 2 1 \rangle = \langle 2 1 | 1 2 \rangle &=& q
\end{eqnarray}

which implies that

\begin{equation}
	-1 \le q = \cos \theta \le 1
\end{equation}

where $\theta$ is the angle between $| 1 2 \rangle$ and $| 2 1 \rangle$.
Therefore for $|q| < 1$, the states $| 1 2 \rangle$ and $| 2 1 \rangle$ are
linearly independent, thus no commutation type of relation between $a^*_{1}$
and $a^*_{2}$ can be formulated.  For $q =1$ the bosonic commutation
relation

\begin{equation}
	a^*_{1} a^*_{2} - a^*_{2} a^*_{1} = 0
\end{equation}

and for $q = -1$ the fermionic commutation relation

\begin{equation}
	a^*_{1} a^*_{2} + a^*_{2} a^*_{1} = 0
\end{equation}

are consistent with the scalar products in (5).

	If we return to the
definition of $N(n,q)$ we see that it can be expanded in terms of
$N(n-1,q)$:

\begin{eqnarray}
	N^{i_1i_2 ...  i_n}_{j_1j_2 ...  j_n} &=&
	N^{i_1}_{j_1} N^{i_2i_3 ...  i_n}_{j_2j_3 ...  j_n} + q
	N^{i_1}_{j_2} N^{i_2i_3 ...  i_n}_{j_1j_3 ...  j_n}
	+ q^2 N^{i_1}_{j_3} N^{i_2i_3i_4 ...  i_n}_{j_1j_2j_4 ...  j_n}
	\nonumber \\ &+& ...  +
	q^{n-1} N^{i_1}_{j_n} N^{i_2i_3 ...  i_n}_{j_1j_2 ...  j_{n-1}}
\end{eqnarray}

where $N^{i_x}_{i_y}$ reduces to the ordinary Kronecker delta,
$\delta_{j_{y}}^{j_{x}}$.  An interesting property is obtained by taking $q
= -1$.  Then $N$ becomes the generalized Kroneckerdelta i.e.,

$$\det [M] = N(n, -1)$$

where the elements of the $n \times n$ matrix $M$ are, $M_{km} =
\delta_{j_{k}}^{j_{m}}$.

	Now we can define a $q$-dependent $\epsilon$ symbol using the
generalized Kronecker delta $N$:

\begin{equation}
	\epsilon_{j_1 j_2 ..  j_n} (q) = N^{12 ...  n}_{j_1j_2 ...
	 j_n}(-q)
\end{equation}

	The above $q$-epsilon symbol is a generalized, q deformed
permutation tensor.  One can see that the new $q$-epsilon tensor reduces to
the ordinary epsilon tensor for $q = -1$.  Writing it down directly in two
dimensions, it is seen that it can be represented by a $2 \times 2$ matrix,

\begin{equation}
	\epsilon_{ij} = \left( \begin{array}{ll} 0 & 1 \\ -q & 0
	 \end{array}\right).
\end{equation}

Note also that $\epsilon^{2} = -q I$, where I denotes the unit matrix.
Another interesting property of the new $q$-epsilon tensor is contraction of
two epsilon tensors over common indices.  Using Einstein convention for
summation, one finds:

\begin{equation}
	\epsilon_{\alpha j_2 j_3 j_4 ...  j_n} \epsilon_{j_2 j_3
	j_4 ...  j_n \beta} = (-q)^{n-1}[n-1]_{q^2}!\delta^{\alpha}_{\beta}
\end{equation}

where $[n]_{q^{2}} = \frac{1-q^{2n}}{1-q^{2}}$ denotes the basic number $n$
with parameter $q^{2}$ \cite{[6]}.  From the definition of $q$-epsilon, the
following very useful index permutation property also follows

\begin{equation}
	\epsilon_{ i ...  j k ...  n} = -q^{-1} \epsilon_{ i ...  k
	j ...  n} \mbox{\quad if \quad}j < k
\end{equation}

	For matrices over the fields of real or complex numbers the group
$SL(n)$ can be defined as the group of matrices which leave the ordinary
Levi-Civita epsilon symbol invariant.  Analogously, we will define the
quantum group $SL_{q}(n)$ as the "group" of matrices with possibly
non-commuting elements which leave the $q$-epsilon symbol invariant.  Thus

\begin{equation}
	a_{i_1j_1} a_{i_2j_2} ...  a_{i_nj_n} \epsilon_{i_1 i_2 ...
	i_n} = \epsilon_{j_1 j_2 ...  j_n}
\end{equation} and

\begin{equation}
	a_{j_1i_1} a_{j_2i_2} ...  a_{j_ni_n} \epsilon_{i_1 i_2 ...
	i_n} = \epsilon_{j_1 j_2 ...  j_n}
\end{equation}

For $|q| \ne 1$, these equations cannot be satisfied if the matrix elements
$a_{ij}$ commute among themselves.  Therefore we assume $a_{ij}$ are
non-commuting elements of a $n \times n$ matrix $A$.

	We will show that for $n=2$ the invertibility of $A$ together with
one of the equations (14) or (15) implies the other.  For $n > 2$, we need
the two equations together because the invertibility of A does not imply the
invertibility of every of its $2\times 2$ submatrices.  By writing (14) in
two dimensions one has,

\begin{equation}
	a_{i_1j_2}a_{i_2j_2}\epsilon_{i_1i_2} = \epsilon_{j_1j_2}
\end{equation}

where $a_{ij} = A_{ij} = \left( \begin{array}{ll} a & b \\ c & d
\end{array}\right)_{ij}$.  (16) contains four equations for two distinct
values of i,j:

\begin{eqnarray}
	a c - q c a &=& 0 \mbox{ \quad (a)} \nonumber \\
	a d - q c b &=& 1 \mbox{ \quad (b)} \nonumber \\
	d a - q^{-1} b c &=& 1 \mbox{ \quad (c)} \nonumber \\
	b d - q d b &=& 0 \mbox{ \quad (d)}
\end{eqnarray}

	At this stage, we define the $q$-determinant of a $2 \times 2$
matrix as in equation (17b), and we note that our matrix $A$ has unit
determinant.  The set of equations (17) are half of the $SL_{q}(2)$ defining
relations \cite{[7]}.  For the other half we return to equation (16) and
write it in matrix form

$$A^T \epsilon A = \epsilon$$

Using the existence of $A^{-1}$ and (11) one obtains,

$$A \epsilon A^{T} = \epsilon$$

which, when solved, gives

\begin{eqnarray}
	a b - q b a &=& 0 \mbox{ \quad (a)} \nonumber \\
	a d - q b c &=& 1 \mbox{ \quad (b)} \nonumber \\
	d a - q^{-1} c b &=& 1 \mbox{ \quad (c)} \nonumber \\
	c d - q d c &=& 0 \mbox{ \quad (d)}
\end{eqnarray}

	Thus in the case of $n=2$, starting from (14) and the invertibility
of A we have obtained the well known conditions for $A \in SL_{q}(2)$. For
the $3$ dimensional case using both (14) and (15), it is possible to obtain
the corresponding result after explicit lengthy calculations. As a byproduct,
$A^{-1}$ in terms of sub $q$-determinants of $A$, is also obtained.

	Let us therefore consider the general $n$ dimensional case and show
 that $SL_{q}(n)$
can be constructed from the invariance of $\epsilon(n,q)$.  At this point
for sake of clarity, we do not use the Einstein convention, instead we will
write all the summations explicitly.  Starting from (15), we can write

\begin{equation}
	\sum\limits_{m_1...  m_{n-2} r s} \epsilon_{m_1 ...
	m_{n-2} r s}a_{i_1m_1} ...  a_{i_{n-2}m_{n-2}}a_{jr}a_{ks} =
	\epsilon_{i_1 ...  i_{n-2} j k}
\end{equation}

\begin{equation}
	\sum\limits_{m_1 ...  m_{n-2} r s} \epsilon_{m_1 ...
	m_{n-2} r s} a_{i_1m_1} ...  a_{i_{n-2}m_{n-2}}a_{kr}a_{js} =
	\epsilon_{i_1 ...  i_{n-2} k j}
\end{equation}

	with $j < k$ .  Using the index permutation property (13), equations
(19) and (20) yield :

\begin{eqnarray}
	\sum\limits_{m_1 ...  m_{n-2} r s} \epsilon_{m_1
	...  m_{n-2} r s} a_{i_1m_1} ...  a_{i_{n-2}m_{n-2}}a_{jr}a_{ks}
	 \nonumber \\
	= -q^{-1}\sum\limits_{m_1 ...  m_{n-2} r s} \epsilon_{m_1 ...
	m_{n-2} r s}a_{i_1m_1} ...  a_{i_{n-2}m_{n-2}} a_{kr}a_{js}
\end{eqnarray}

We can divide the summation over r,s into two parts such that

$$\sum\limits_{r,s} = \sum\limits_{r<s}+\sum\limits_{r>s}$$ since
 $\epsilon_{m_{1} ...  m_{n-2}rr}$ is zero.

Now, renaming the indices in the $r > s$ parts of the summations in (21), we
obtain

\begin{eqnarray}
	\sum\limits_{m_1 ...  m_{n-2} r<s} \epsilon_{m_1
	...  m_{n-2} r s} a_{i_1m_1}...a_{i_{n-2}m_{n-2}}a_{jr}a_{ks}
	\nonumber \\
     + \sum\limits_{m_1 ...  m_{n-2} s>r} \epsilon_{m_1 ...  m_{n-2}sr}
	a_{i_1m_1}...a_{i_{n-2}m_{n-2}}a_{js}a_{kr} \nonumber \\
     = -q^{-1}\biggl(\sum\limits_{m_1 ...  m_{n-2} r<s}\epsilon_{m_1 ...
      m_{n-2} r s} a_{i_1m_1}...a_{i_{n-2}m_{n-2}}a_{kr}a_{js} \nonumber \\
     + \sum\limits_{m_1 ...  m_{n-2} s>r} \epsilon_{m_1 ...  m_{n-2} s
	r}a_{i_1m_1}...a_{i_{n-2}m_{n-2}}a_{ks}a_{jr}\biggr)
\end{eqnarray}

If $\epsilon_{m_{1}...  m_{n-2}rs}$ is factored out using the index
permutation property, (22) reads:

\begin{eqnarray}
	\sum\limits_{m_1 ...  m_{n-2} r<s}
	a_{i_1m_1}...a_{i_{n-2}m_{n-2}} \epsilon_{m_1 ...  m_{n-2} r
	s}\nonumber \\
	 \left(\left(a_{jr}a_{ks}-qa_{js}a_{kr}\right)+q^{-1}
	\left(a_{kr}a_{js}-qa_{ks}a_{jr}\right)\right)  =0
\end{eqnarray}

$A$ being invertible, matrix multiplying this equation from the left gives

\begin{equation}
 	\sum\limits_{r<s} \epsilon_{m_1 ...  m_{n-2} r s}
	\left(\left(a_{jr}a_{ks}-qa_{js}a_{kr}\right)+q^{-1}
	\left(a_{kr}a_{js}-qa_{ks}a_{jr}\right)\right)=0
\end{equation}

Noting that the $n-2$ indices $m_{i}$ in the above epsilon tensor are free,
any choice of them such that epsilon is non vanishing uniquely determines
both $r$ and $s$, since $r < s$.  Noting also that in the passage from (20)
to (21) $j < k$, the contents of (24) can be rewritten as

\begin{equation}
	\left(a_{jr}a_{ks}-qa_{js}a_{kr}\right)+q^{-1}
	\left(a_{kr}a_{js}-qa_{ks}a_{jr}\right)=0\mbox{\quad with \quad}
	r<s, j<k
\end{equation}

The scheme below will assist in visualizing (25).  Define an arbitrary 2x2
submatrix of $A$ as:

$$A = \left( \begin{array}{ll} a & b \\ c & d \end{array}\right) = \left(
	\begin{array}{ll} a_{jr} & a_{js} \\ a_{kr} & a_{ks}
	\end{array}\right)$$

and use (25) to obtain

$$ad - qbc = da - q^{-1}cb$$

If we start from (14) instead of (15), using similar steps, but matrix
multiplying from right by $A$ instead of left we can get

\begin{equation}
	a_{jr} a_{ks} - q a_{kr} a_{js} = a_{ks} a{jr} -q^{-1}
	a_{js} a_{kr}\mbox{\quad with \quad} r<s, j<k
\end{equation}

This equation when applied to our arbitrary $2 \times 2$ submatrix gives:

$$ad - qcb = da - q^{-1}bc$$

If we combine (25) and (26), we obtain

\begin{equation}
	a_{js} a_{kr} = a_{kr} a_{js}
\end{equation}

which is valid for every $2 \times 2$ submatrix of $A$.  (27) Gives for our
arbitrary submatrix:

$$bc = cb$$

	Up to now we have derived the relations between diagonal elements of
every $2 \times 2$ submatrix of A.  For the relations between row/column
elements we start from (19) and set $j=k$, so that the right hand side
vanishes.  Then as in the previous case, we divide the equation into two
parts:

\begin{eqnarray*}
	\sum\limits_{m_1 ...  m_{n-2} r<s} \epsilon_{m_1 ... m_{n-2} r
	s}a_{i_1m_1} ...  a_{i_{n-2}m_{n-2}}a_{kr}a_{ks}\\ +
	\sum\limits_{m_1 ...  m_{n-2} r>s} \epsilon_{m_1 ... m_{n-2} r
	s}a_{i_1m_1} ...  a_{i_{n-2}m_{n-2}} a_{kr}a_{ks} = 0
\end{eqnarray*}

Renaming the indices of the second part and using the index permutation
property to collect similar terms, one obtains:

\begin{equation}
	\sum\limits_{m_1 ...  m_{n-2} r<s}a_{i_1m_1} ...
	a_{i_{n-2}m_{n-2}} \epsilon_{m_1 ...  m_{n-2} r
	s}\left(a_{kr}a_{ks}-qa_{ks}a_{kr}\right)=0
\end{equation}

Using the invertibility of $A$ and matrix multiplying from left implies:

\begin{equation}
	a_{kr} a_{ks} - q a_{ks} a_{kr} = 0
\end{equation}

since for any fixed set of indices m such that epsilon is non-zero , $r$ and
$s$ are also fixed as in the previous case.  The above equation will give us
(18a,b) for any $2 \times 2$ submatrix of $A$.  The "transpose" of the
equation (19), treated in a similar way will yield the rest of the
$GL_{q}(n)$ definition relations, namely (17a,b).  Thus every submatrix of
$A \in GLq(2)$.  It follows that $A \in SLq(n)$.

	We have thus constructed a $q$-epsilon tensor from certain inner
products of states of a multi-dimensional $q$-oscillator.  We should point
out however, that this construction is not unique in the sense that there
exist other $n$-dimensional $q$- oscillators that realize the same epsilon
tensor.  For example in the case of Pusz- Woronowicz (PW) oscillators
\cite{[3],[9]}, the (same) epsilon tensor is realized via (4),(10) with
the creation and annihilation operators exchanged by the corresponding
$PW$ operators. However, the algebra of the latter is much more
complicated.  We will touch upon this shortly.

	On the other hand in terms of the Manin quantum
hyperplane \cite{[8]}, equations (14) and (15) have a very natural
interpretation.  We recall that the quantum hyperplane is spanned by
non-commuting coordinates, derivatives and differentials, $x^{i}$,
$\partial_{i}$ and $\xi^{i}$, repectively.  In particular the commutation
relations among the differentials are

\begin{equation}
	\xi^i \xi^j = -q^{-1} \xi^j \xi^i \mbox{\quad for \quad} i>j
\end{equation}

and remain invariant under a transformation $A \in SL_{q}(n)$, $\xi^{i}
\longrightarrow a^{i}_{j}\xi^{j}$.  (30) is very reminiscent of the index
permutation property (13).  In fact the expression

\begin{equation} \xi^i_1 \xi^i_2 ...  \xi^i_n = \epsilon_{i_1 i_2 ...  i_n}
		\xi^1 \xi^2 ...  \xi^n
\end{equation}

is $SL_{q}(n)$ covariant by virtue of either (14) or (15), analogously to
the classical, i.e.  non-deformed case.  The only additional condition
needed to incorpoprate the information content of both (14) and (15) is the
statement

\begin{equation}
	det_q A = det_q\left(A^T\right)
\end{equation}

which is of course trivial in the non-deformed case.

	The relations between
the $q$-oscillator (1) and quantum gropus have been obtained rather
indirectly using the $q$-epsilon tensor.  However it turns out that
there are yet more relations between the $q$-oscillators (1) and
other $q$-oscillators related to quantum groups.  Consider, again,
the Pusz-Woronowicz oscillators \cite{[3],[9]}.  Denoting states
created from the vacuum by a subindex PW, one can write down an
explicit formula relating these to the states of the oscillator (1),

\begin{equation}
	|i_{1} i_{2} ...  i_{n}\rangle_{PW} = |i_{1}' i_{2}' ...
	i_{n}'\rangle\langle i_{1}' i_{2}' ...  i_{n}'|i_{1} i_{2} ...
	i_{n}\rangle C_{i_{1} i_{2} ...  i_{n}}
\end{equation}

\begin{equation}
	C_{i_{1} i_{2} ...  i_{n}} = \sqrt{\frac{[r_1]_{q^2}
	!}{\left([r_1]_{q}!\right)^3} \cdot \frac{[r_2]_{q^2}
	!}{\left([r_2]_{q}!\right)^3}\cdot ...  \cdot\frac{[r_m]_{q^2}
	!}{\left([r_m]_{q}!\right)^3}}\mbox{\quad with \quad} m\le n
\end{equation}

where $r_{i}$ is the number of repeated indices among $i_{1} i_{2} ...
i_{n}$, $i_{1}' i_{2}' ...  i_{n}'$, denotes the ordering of $i_{1} i_{2}
...  i_{n}$ such that $i_{1}'\le i_{2}'\le ...  \le i_{n}$' and $|i_{1}
i_{2} ...  i_{n}\rangle_{PW}$ are the states created by the PW creation
operator.  Note that in (33) all states on the RHS belong to the oscillator
(1).

	Perhaps the simplest multidimensional $q$-oscillator is the one
which consists of D commuting copies of the one dimensional $q$-oscillator.
This oscillator is related to the $PW$ oscillator by a transformation
\cite{[10]}.  It is also related to the Coon- Baker-Yu oscillator (1), since
its $n$-particle states are just the fully symmetrized $n$-particle states
of the CBY oscillator.

	To conclude, we have arrived at a construction of
$SL_{q}(n)$ via the $q$-epsilon tensor.  It turns out that the $U(n)$
invariant CBY oscillator (1) is closely related to this, in the sense that
the corresponding algebra possesses the minimal structure necessary to
realize this $q$-epsilon tensor via innerproduct states, (4),(10).

\end{document}